\documentclass[twocolumn]{aastex631}
\usepackage{bm}
\usepackage{amsmath}
\newcommand{\Rsun}{R_\odot}
\newcommand{\kpar}{\kappa_\parallel}
\newcommand{\lpar}{\lambda_\parallel}

\shorttitle{Does DSA reach steady state?}
\shortauthors{Jebaraj et al.}

\begin{document}

\title{Does Diffusive Shock Acceleration Reach Steady State? \\ Insights from a Near-Sun Shock Observed by \textit{Parker Solar Probe}}

\author[0000-0002-0606-7172]{I. C. Jebaraj}
\affiliation{Department of Physics and Astronomy, University of Turku, FI-20014 Turun yliopisto, Finland}

\author[0000-0001-6360-1987]{M.~A.~Malkov}
\affiliation{Department of Astronomy and Astrophysics, University of California, San Diego, La Jolla, CA 92093, USA}
\affiliation{Eureka Sci., Oakland, CA, USA}

\author[0000-0001-6344-6956]{N.~Wijsen}
\affiliation{Centre for mathematical Plasma Astrophysics, KU Leuven, 3001 Leuven, Belgium}

\author[0000-0001-6427-1596]{O.~V.~Agapitov}
\affiliation{Space Sciences Laboratory, University of California, Berkeley, CA 94720, USA}
\affiliation{Astronomy and Space Physics Department, National Taras Shevchenko University of Kyiv, 03127 Kyiv, Ukraine}

\author[0000-0001-6589-4509]{A.~Kouloumvakos}
\affiliation{The Johns Hopkins University Applied Physics Laboratory, Laurel, MD 20723, USA}

\author[0000-0001-9325-6758]{A. Afanasiev}
\affiliation{Department of Physics and Astronomy, University of Turku, FI-20014 Turun yliopisto, Finland}

\author[0000-0002-5456-4771]{F.~Fraschetti}
\affiliation{Center for Astrophysics $|$ Harvard $\&$ Smithsonian, Cambridge, MA, 02138, USA}

\author[0000-0002-1349-3663]{E.~Husidic}
\affiliation{Department of Physics and Astronomy, University of Turku, FI-20014 Turun yliopisto, Finland}

\author[0000-0002-0978-8127]{C. M. S. Cohen}
\affiliation{California Institute of Technology, 1260 E California Blvd, Pasadena, CA 91125, USA}

\author[0000-0002-3298-2067]{R. Vainio}
\affiliation{Department of Physics and Astronomy, University of Turku, FI-20014 Turun yliopisto, Finland}

\begin{abstract}
Collisionless shocks accelerate particles at a rate set by the diffusion coefficient of those particles in waves they themselves generate. Theory fixes that coefficient only in steady state, and no measurement has tested whether an evolving shock ever reaches that state. We report \textit{Parker Solar Probe} observations of a fast, near-parallel interplanetary shock at 0.24~au, encountered four hours after eruption, with the upstream waves and the protons they scatter recorded on the same magnetic field line. From the wave spectra alone we derive the parallel mean free path of 0.05--30~MeV protons, resolved in energy and in distance from the shock. The mirror force of the measured compressive fluctuations supplies the scattering through the pitch angles that gyroresonance cannot reach. The mean free path grows with distance nearly as the steady self-generated solution requires, yet the steady balance holds at one energy only, near 1.3~MeV. The intensity profiles are a factor of two shallower than the measured coefficient predicts, because the waves are still growing. Confinement fails between 2.3 and 6~MeV, beginning a factor of two below what the age of the shock allows. The growth of the waves its own particles drive, rather than its age or its size, limits the energy an evolving shock confines.
\end{abstract}


\section{Introduction}
\label{sec:intro}

Diffusive shock acceleration (DSA) is the standard theory for cosmic-ray acceleration at collisionless shocks \citep{Krymskii1977,Axford1977,Bell78,Blandford1978}. In DSA, particles gain energy through repeated shock crossings, enabled by pitch-angle scattering from magnetic fluctuations, including waves generated upstream by the accelerated particles themselves \citep{Bell78,Lee83}. Standard DSA assumes those waves have finished growing (a quasi-steady scattering environment) at all rigidities confined by the shock \citep{MalkovDrury01}. At a fast, evolving shock, however, particle acceleration and wave growth can occur on comparable timescales. Simulations with self-generated waves reach that steady state when their resonance condition ignores the pitch angle \citep{VainioLaitinen07,Vainio14} and not when it retains it \citep{Afanasiev15}, and no in situ measurement has established whether the system ever reaches it.

The particle energy a shock can reach has three limits. \citet{Lagage83} set the first limit, the time to complete enough acceleration cycles, $t_{\rm acc}\propto\kpar/u_1^2$ with $\kpar$ the parallel diffusion coefficient and $u_1$ the upstream flow speed in the shock frame. They assumed that the induced turbulence does not substantially exceed the magnitude of the ambient field. The second is the size of the accelerator, which the diffusion length $\kpar/u_1$ must not exceed, and at heliospheric shocks adiabatic focusing in the diverging field can set that size \citep{Vainio14}. \citet{Bell04} showed that the streaming particles amplify the confining field far beyond its ambient value, which raises the maximum energy allowed by both limits. \citet{Bell13} added a third limit by requiring the amplified field to grow within the lifetime of the source. This growth-time limit is the only one of the three that does not follow from $\kpar$ alone.

\citet{Vainio03} made the third limit quantitative for a traveling source and showed that the proton streaming through a plasma element, integrated over time rather than taken at its instantaneous value, controls the wave amplification the element experiences. The precursor traps particles only while that integral is large enough for the resonant waves to grow. The rigidity at which that condition fails is then the confinement boundary, and the highest rigidity this boundary reaches over a shock's history sets the maximum energy of the event. \citet{Lee05} located the same boundary through the self-consistent ion escape rate, and in self-generated-wave simulations it advances as the waves grow \citep{VainioLaitinen07,NgReames08,Afanasiev23}. Particles above it do not return to the front \citep{Do25}, and the upstream intensity steepens within one diffusion length of the shock \citep{Fraschetti21}. No study has evaluated the two limits that follow from $\kpar$ with an in situ measurement of the diffusion coefficient at a shock, or compared the three at the same shock.

In situ measurements of the diffusion coefficient have so far sampled only particles the shock still confines. At Earth's bow shock the diffuse-ion intensity falls exponentially with upstream distance \citep{Ipavich81,Trattner94}, and multi-spacecraft gradients have yielded diffusion coefficients for tens-of-keV ions \citep{Kis04,Kis18}. \citet{Kennel86} tested the coupled wave and particle predictions of \citet{Lee83} against International Sun-Earth Explorer (ISEE) observations of 35--200~keV ions at an interplanetary shock. The particle profiles agreed, but the resonant wave spectrum did not. Closer to the Sun, \citet{Chen24} computed quasi-linear \citep[QL;][]{vedenov1961nonlinear} diffusion coefficients from the ambient solar wind that \textit{Parker Solar Probe} \citep[PSP;][]{Fox2016} measures, and \citet{Cuesta25} compared such coefficients with those fitted to the upstream intensity profiles of eight PSP shocks below 600~keV. None of these measurements has resolved the diffusion coefficient of a single shock as a function of both energy and distance from the shock. Recent arrival-time studies find events in which the particles above about 1~MeV arrive in inverse order, the higher energies later, ahead of an approaching shock. They attribute the delay to acceleration still in progress or to evolving source--observer connectivity, an interpretation not yet tested against simultaneous wave measurements \citep{Chen25,Kouloumvakos25,Ding25,Li25,Kouloumvakos26b,Allen26,Xu26}.

PSP crossed the shock of 2023 March 13 at 0.24~au, four hours after the eruption that drove it. The shock's upstream speed in the shock frame was $u_1\simeq2400$~km\,s$^{-1}$, its obliquity $\theta_{Bn}\simeq8\arcdeg$ \citep{Jebaraj24,Jebaraj24synchrotron,Wijsen25}, and its Alfv\'enic Mach number $M_A\simeq7$--$7.5$ using the upstream Alfv\'en speed reported in Paper~I \citep{Jebaraj26}. PSP recorded the wave field and the scattered protons on the same field line.

Paper~I resolved the upstream wave field into right- and left-handed branches, and into a field-aligned Alfv\'enic component and an oblique compressive fast-magnetosonic population. \citet{Kouloumvakos26} (Paper~II) tracked the spectral folding energy, at which each evolving upstream spectrum turns over from its low-energy plateau into its power law, and inferred from the evolution of that energy the amplification time of the confining waves. Paper~II also located the transition to escape at 2--5~MeV in the spacecraft frame from the reversal of the downstream anisotropy. Here we derive $\kpar$ from the waves channel by channel, test it against the particle gradients, and compare the three limits at a single shock. We give every step from the field measurements to the transport coefficients in Appendix~\ref{app:method}.

\section{The mean free path from the wave field}
\label{sec:coeff}

\begin{figure}[!t]
  \centering
  \includegraphics[width=\columnwidth]{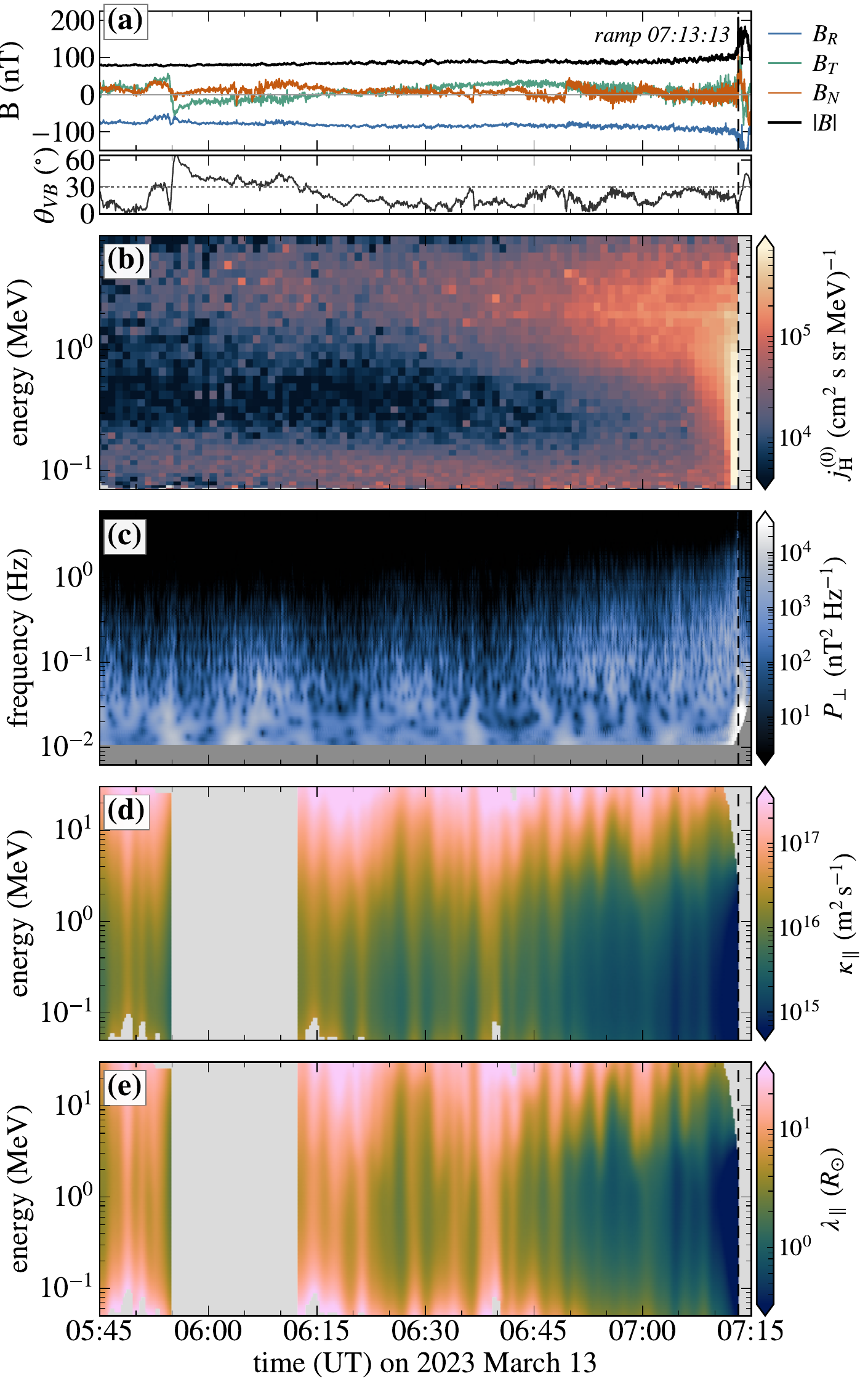}
  \caption{The precursor over 05:45--07:15~UT on 2023 March 13, with the ramp at 07:13:13~UT dashed. \textit{(a)} Magnetic field in radial, tangential, and normal (RTN) components and magnitude, 1~s means, with the field-to-flow angle $\theta_{VB}$ beneath and the dotted line at $30\arcdeg$. \textit{(b)} Isotropic proton intensity $j^{(0)}_{\rm H}(E,t)$. \textit{(c)} Transverse magnetic wavelet power $P_\perp(f,t)$ against spacecraft-frame frequency. \textit{(d)} the parallel diffusion coefficient $\kpar(E,t)$ and \textit{(e)} the parallel mean free path $\lpar(E,t)$ in 15~s bins. Gray marks excluded bins: the post-ramp interval, zero intensity in (b), the cone of influence in (c), and rejected estimates in (d) and (e).}
  \label{fig:shock}
\end{figure}

\subsection{Protons and the waves they drive}

Figure~\ref{fig:shock}a shows the magnetic field from the FIELDS fluxgate magnetometer \citep[MAG;][]{Bale16} through the precursor. The background field $B_0$ rises from $\simeq87$~nT in the outer precursor to $\simeq98$~nT in the last half solar radius. The field stays within $30\arcdeg$ of the flow direction that the wave analysis requires, except during 05:55--06:12~UT (Appendix~\ref{app:method}). The spacecraft-potential electron density $n_e$ rises from $\simeq31$ to $\simeq38$~cm$^{-3}$, nearly in proportion, so the Alfv\'en speed $v_A=B_0/\sqrt{\mu_0 n_e m_p}$ stays at $340$~km\,s$^{-1}$ across the precursor.

The isotropic proton intensity in Figure~\ref{fig:shock}b is the monopole $j^{(0)}_{\rm H}$ of the proton distribution, written $j$ below, measured by the EPI-Lo instrument of the Integrated Science Investigation of the Sun \citep[IS$\odot$IS;][]{McComas16,Hill17}. Above 1~MeV the protons fill the precursor throughout the record. Between 0.2 and 1~MeV they form a suppressed band whose intensity lies a factor $\simeq$5 below the 3~MeV intensity through the middle of the precursor, 06:20--06:40~UT, and fills in only during the last few minutes. From that interval to the final pre-ramp minute the 0.3--0.5~MeV intensities rise by factors of 140--200, while the 3~MeV intensity rises only sixfold.

These protons excite the transverse wave power of Figure~\ref{fig:shock}c. It is broadband and intensifies by an order of magnitude toward the shock across the frequencies with which 0.05--5~MeV protons gyroresonate ($\simeq$0.03--0.4~Hz). The two-component test of \citet{Bieber96}, evaluated in the $\bm{B}_0$--flow plane, assigns a median slab fraction of 0.83 to the transverse power, a share of the reduced power rather than of the fluctuation energy (Appendix~\ref{app:method}). The measured phase speed, the flow past the spacecraft plus the Alfv\'en speed, converts spacecraft-frame frequency to parallel wavenumber.

We measure distances ahead of the shock along its normal as $x=u_1(t_{\rm shock}-t)$, with $t_{\rm shock}$ the time of the ramp. The front approaches the spacecraft at the closing speed $U_0=u_1+V_n\simeq2830$~km\,s$^{-1}$, with $V_n\simeq430$~km\,s$^{-1}$ the upstream flow along the normal in the spacecraft frame, the heliocentric wind plus the sunward motion of PSP. In this coordinate $U_0$ enters every steady balance and shock-frame quantities use $u_1$ (Appendix~\ref{app:profiles}). The true shock-normal distance is $x_n=(U_0/u_1)\,x=1.18\,x$. 

The energetic-particle pressure falls upstream with the $e$-folding scale $L_{\rm EP}=5.5\,\Rsun$ (Paper~I), so a plasma element spends the residence time $\tau_{\rm c}=L_{\rm EP}/u_1\simeq26$~min in the precursor. We divide the precursor at $3\,\Rsun$ into an inner and an outer region. The inner region begins at $0.1\,\Rsun$ for the 15~s region statistics and at $0.2\,\Rsun$ for the 60~s particle profiles, which resolve no profile inside that distance. The outer region ends at $12\,\Rsun$, or $13\,\Rsun$ for the $\lpar(x)$ statistics. All numbers below are medians over the 15~s intervals, written $a\,[b,c]$, with $[b,c]$ denoting the 16th--84th percentile range.

\begin{figure*}[!t]
  \centering
  \includegraphics[width=\textwidth]{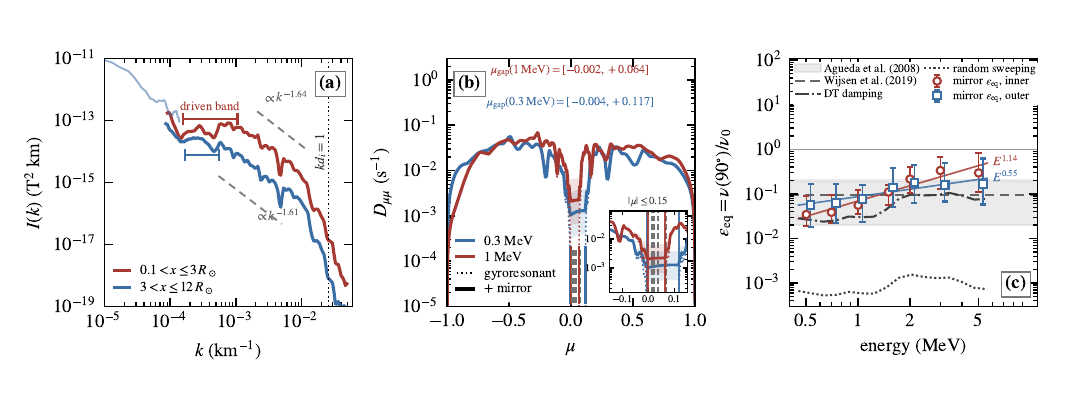}
  \caption{The slab spectrum, the pitch-angle diffusion coefficient, and the equivalent $90\arcdeg$ constant. \textit{(a)} Slab spectrum $I(k)$, the median over the inner (red) and outer (blue) precursor, with the Welch spectrum of the $3$--$20\,\Rsun$ record (light blue). Bars mark the driven bands, dashed lines the fitted falls beyond them, and the dotted line $kd_i=1$. \textit{(b)} $D_{\mu\mu}$ at 0.3~MeV (blue) and 1~MeV (red) on the inner median field, gyroresonance alone (dotted) and with the mirror interaction at $\eta_{\rm TTD}=1$ (solid), shaded over $\eta_{\rm TTD}=0.3$--$3$. Ticks along the bottom axis mark the gyroresonant edges, solid for the measured branches and gray for a non-dispersive field of the same bandwidth, and the inset enlarges $|\mu|\leq0.15$. \textit{(c)} $\epsilon_{\rm eq}$ per region, medians over the 15~s samples with 16th--84th bars, circles inner and squares outer, with fitted slopes at the right. The dynamical-turbulence damping (DT) and random-sweeping curves use the inner median field. The band and the dashed line mark the constants of \citet{Agueda08} and \citet{Wijsen19} at $\epsilon_{\rm eq}=2\epsilon$, and the thin line $\epsilon_{\rm eq}=1$.}
  \label{fig:gap}
\end{figure*}

\subsection{Mirror scattering across the dispersive gap}

Gyroresonance with the two transverse branches and the mirror force of the compressive family set the pitch-angle diffusion coefficient $D_{\mu\mu}$. Protons of pitch-angle cosine $\mu$, measured in the solar-wind frame along the anti-sunward direction the branches propagate in, gyroresonate with the right-handed branch through the $n=-1$ harmonic and with the left-handed branch through $n=+1$. The measured reduced helicity $\sigma_m(k)$ partitions the one-sided slab spectrum $I(k)$ between the two branches (Figure~\ref{fig:gap}a). We evaluate the resonance on the cold-plasma dispersion relations $\omega_R(k)$ and $\omega_L(k)$ rather than in the magnetostatic limit of \citet{Jokipii66}, since at the top of the measured band $\omega_R$ reaches $4.5\,\Omega$, with $\Omega$ the proton gyrofrequency. We broaden each resonance while preserving its integrated area \citep{Volk73,Volk75,Xu18}, using the random-sweeping model of \citet{Bieber94} with its decorrelation parameter set by the measured slab amplitude $\delta B_\perp/B_0$.

The dispersion widens the pitch-angle gap that magnetostatic QL theory leaves around $90\arcdeg$ and shifts it further toward positive $\mu$ (Figure~\ref{fig:gap}b). On the whistler side the rising $\omega_R$ raises the resonant pitch-angle cosine. Below a parallel speed of $(3\sqrt{3}/2)\,v_A=2.60\,v_A$, the tangency condition of the $n=-1$ whistler resonance \citep{Vainio00}, the resonance leaves the branch, at $\mu=+0.064$ at 1~MeV (inset). The tangency occurs at $kd_i=2/\sqrt{3}$, with $d_i$ the ion inertial length, inside the measured band, so the dispersion relation rather than the bandwidth sets this edge. On the ion-cyclotron side the resonance reaches $\mu=-0.002$ on the cold branch and $\mu=-0.014$ when kinetic damping truncates the branch at $kd_i\gtrsim0.8$.

A non-dispersive Alfv\'enic field of the same bandwidth would leave the gap $|\mu-v_A/v|<\Omega/(vk_{\rm max})=0.013$, with $v$ the proton speed and $k_{\rm max}$ the top of the measured band. At 1~MeV this gap spans $[+0.012,+0.038]$. The Alfv\'enic phase speed therefore produces most of the displacement of the measured gap, with $v_A/v=0.025$ against a measured gap center of $0.031$, whereas dispersion widens the gap by a factor of 2.6. A Galilean boost along $\bm{B}$ shifts the gap but leaves the width in parallel speed unchanged.

The mirror force of the compressive family crosses that gap through transit-time damping (TTD), the $n=0$ Cherenkov resonance at its parallel phase speed, $v\mu=c_f(\theta_{kB})/\cos\theta_{kB}$ \citep{Achterberg81,Schlickeiser98,YanLazarian04}. Here $c_f$ is the fast-magnetosonic phase speed, evaluated in each interval at the median wave normal angle $\theta_{kB}=61\arcdeg$ of that family (Paper~I), and the median over the intervals of the resonant speed is 740~km\,s$^{-1}$ across the precursor. The measured compressive power fixes the amplitude of the mirror force.

In the adiabatic mirror approximation a compression reflects any proton whose parallel speed lies within $v_\perp(\delta B_\parallel/B_0)^{1/2}$ of the resonant speed, with $v_\perp$ the perpendicular proton speed and $\delta B_\parallel$ the compressive amplitude, $0.012\,B_0$ in the outer precursor and $0.022\,B_0$ in the inner. At 1~MeV the window extends 1500--2000~km\,s$^{-1}$ to either side of the resonant speed in the solar-wind frame and spans the gyroresonant gap, fixed in parallel speed from $-30$ to $+880$~km\,s$^{-1}$, so the mirror force moves protons between its whistler and ion-cyclotron sides. Gyroresonant scattering on either side of the gap removes a proton from the window at 1~MeV after a median of 2~s in the inner precursor and 7~s in the outer, under a quarter of the bounce period, so few reflections complete. The sense of propagation, which Paper~I leaves open, sets the center of the window (Appendix~\ref{app:mc}). The family's own cyclotron resonance lies above 100~MeV (Paper~I), so only its mirror force acts on these protons. The one factor in $D_{\mu\mu}$ that no measurement fixes is the mirror efficiency $\eta_{\rm TTD}$, of order unity, which we vary over 0.3--3. Simulating this event with the mirror interaction would constrain it.

\begin{figure*}[!t]
  \centering
  \includegraphics[width=\textwidth]{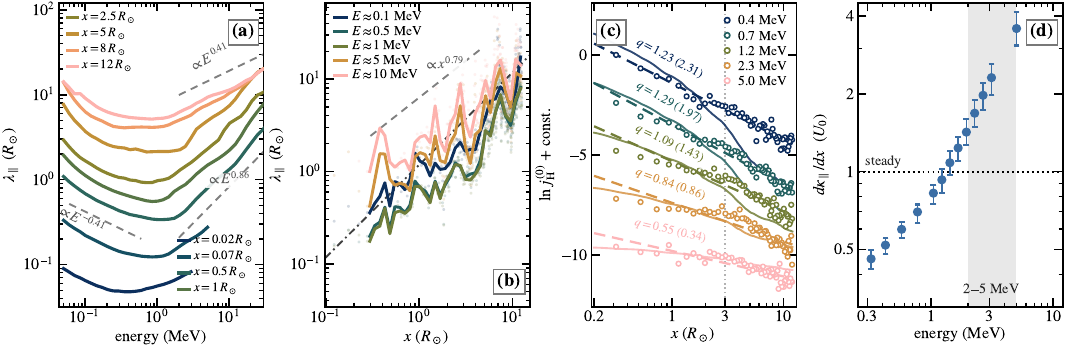}
  \caption{The parallel mean free path against energy and against distance, the intensity profiles, and the growth of the diffusion coefficient. \textit{(a)} $\lpar(E)$ at the labeled distances, medians over the 15~s bins near each, single bins at $0.02$ and $0.07\,\Rsun$, with the fitted slopes as dashed lines. \textit{(b)} $\lpar(x)$ at the labeled energies, medians in logarithmic distance bins over the 15~s values (faint dots), with the joint power-law slope dashed and $\lpar=x_n$ dot-dashed. \textit{(c)} Isotropic intensity in five channels (circles, offset vertically), the power-law fits $j\propto x^{-q}$ (dashed) labeled by $q$, and the steady profile on the measured $\kpar(x)$ of each channel (solid) with its fitted index in parentheses. The dotted line marks $3\,\Rsun$. \textit{(d)} Fitted growth speed $d\kpar/dx$ in units of $U_0$ with 16th--84th percentile bars, the dotted line at $U_0$ and the gray band the 2--5~MeV transition of Paper~II in the spacecraft frame.}
  \label{fig:slices}
\end{figure*}

\subsection{A minimum near 1~MeV and growth with distance}

The parallel mean free path $\lpar$ is the pitch-angle integral of $D_{\mu\mu}$ \citep{HasselmannWibberenz70}, evaluated for energies from 0.05 to 30~MeV, with $\kpar=v\lpar/3$ and the effective scattering frequency $\nu_{\rm eff}=v/\lpar$. Above $\simeq$1~MeV the mean free path grows with energy, and at every energy it shortens toward the shock (Figure~\ref{fig:shock}d,e). At 1~MeV its median over the precursor is $2.7\,[0.7,6.2]\,\Rsun$. The product of $L_{\rm EP}$ with the closing speed gives a benchmark independent of the scattering model, $\kappa_{\rm eff}=U_0L_{\rm EP}=1.1\times10^{16}$~m$^2$\,s$^{-1}$, and the precursor median of $\kpar$ at 1~MeV is 0.79 of this value. The ambient-wind coefficient at this distance and energy \citep{Chen24} is 1.2 times this benchmark and twenty times the wave value over the last half solar radius.

In the last half solar radius before the ramp the mean free path at 1~MeV drops to $0.21\,[0.14,0.25]\,\Rsun$, $\simeq10^2$ proton gyroradii $r_g$, and the effective scattering frequency peaks at $\nu_{\rm eff}\simeq0.10$~s$^{-1}$ near 1.6~MeV. Below $\simeq$0.3~MeV the scattering through $90\arcdeg$ controls the mean free path and the streaming anisotropy $3u_1/v$ exceeds unity, so the transport is diffusive only above that energy. Across 0.3--2.6~MeV the medians over the last half solar radius span $0.20$--$0.34\,\Rsun$. A shallow minimum persists at every distance in Figure~\ref{fig:slices}a, lying between 0.7 and 1~MeV through most of the precursor and shifting to 0.4~MeV at the innermost bins, down to the final pre-ramp bin at $x=0.02\,\Rsun$.

The mirror interaction that crosses the dispersive gap makes that minimum. Dispersion raises $\lpar$ in the inner precursor above its non-dispersive value by 9 percent at the minimum and by 50 percent at 0.05~MeV. Toward low energy the whistler branch outruns the particles and the resonance retreats to $\mu\geq2.6\,v_A/v$. The pitch angles where the mirror dominates supply 84 percent of the $\lpar$ integral at 0.05~MeV in the inner precursor, compared with 10 percent at 5~MeV. The mean free path therefore increases toward lower energy, as $E^{-0.41}$ over 0.06--0.6~MeV at $x=0.5\,\Rsun$. The exponent runs from $-0.51$ to $-0.29$ over $\eta_{\rm TTD}=0.3$--3 and from $-0.32$ to $-0.48$ across the three dispersion treatments (Appendix~\ref{app:method}). Toward high energy the gyroresonances recede into the driven band, the wavenumber interval over which the spectrum flattens against its surroundings, and $\lpar$ rises again. The self-generated-wave simulations of \citet{Afanasiev15,Afanasiev23} fill the resonance gap with a prescribed rate, and their mean free path grows with energy throughout, without this minimum.

A power law fitted to $I(k)$ across that band gives $k^{+0.3}$ in the inner precursor and $k^{-0.9}$ in the outer, against falls of $k^{-1.64\pm0.09}$ and $k^{-1.61\pm0.06}$ beyond the band, both within one sigma of $k^{-5/3}$. In the outer precursor the band spans the $|\mu|=1$ resonances of 1.2--13~MeV protons and in the inner it widens to 0.36--16~MeV (Figure~\ref{fig:gap}a). As the intensity of lower-energy protons rises toward the shock, the growth extends to their resonances. In the magnetostatic scaling $\lpar\propto E^{(2-s)/2}$, with $s$ the spectral index of $I(k)$, a flat band gives $\lpar\propto E$. The near-shock rise is $E^{0.86}$ over 2--30~MeV at $x=0.5\,\Rsun$, against the $E^{0.2}$ that a pure power law of the same integrated power would give.

The growth of $\lpar$ with distance is $x^{0.79\pm0.05}$ jointly across the five channels of Figure~\ref{fig:slices}b over $0.3$--$13\,\Rsun$, and $x^{0.85\pm0.06}$, $x^{0.90\pm0.06}$, and $x^{0.86\pm0.07}$ at the 0.4, 1, and 2.6~MeV channels of the profile analysis on the same distance bins. The steady self-generated solution is $\lpar=3(u_1-v_A)\,x_n/v$ \citep{Bell78,VainioLaitinen07}, linear in distance. The measured growth is shallower and the amplitude agrees. The power-law fit to the median-binned $\lpar(x)$ gives $0.52\,\Rsun$ at $x=1\,\Rsun$ and 1~MeV against $0.53\,\Rsun$ from the steady solution, within the 20--25 percent that the shock-speed uncertainty allows. The amplitude depends on the resonance broadening, and the mirroring-broadened width of \citet{Volk75} and \citet{Xu18} applied to the same spectra gives $\simeq0.2\,\Rsun$ instead, an upper limit on the change (Appendix~\ref{app:mc}).

The steady solution falls as $E^{-1/2}$ at every distance, whereas far upstream the measured $\lpar$ rises as $E^{0.41}$ over 2--30~MeV at $x=12\,\Rsun$. At the top of the range the rise steepens toward the shock, as the resonances of the fastest protons cross the driven band toward its low-wavenumber edge. Nearest the ramp the cone of influence truncates that end above 3.5~MeV, and we reject those estimates (Appendix~\ref{app:method}).

The band-integrated slab amplitude rises from $\delta B_\perp/B_0\simeq0.07$ beyond $3\,\Rsun$ to $0.14$ between 0.5 and $3\,\Rsun$, and to $0.26$ inside $0.5\,\Rsun$. The vector root-mean-square (rms) amplitude reaches $\delta B/B_0\simeq0.6$ in the final 15~s, with individual excursions beyond unity, and \citet{Jebaraj24} measure $\delta B/B_0\gtrsim0.5$ across the shock layer. The rms amplitude implies a Bohm limit $\lpar/r_g\sim(B_0/\delta B)^2\simeq3$, a factor of 36 below the $\simeq10^2\,r_g$ derived over the last half solar radius, so the transport in that layer lies outside the QL description. Particles above 1~MeV have shock-frame diffusion lengths $\kpar/u_1\gtrsim0.4\,\Rsun$, eight times the $0.05\,\Rsun$ the final 15~s layer spans, so the precursor coefficient sets their acceleration.

\subsection{The measured scattering through 90 degrees}

The transverse spectrum does not fix the scattering through $90\arcdeg$. Transport calculations fill the gap with an empirical parameter, written $H$ or $\epsilon$, for the nonlinear corrections near $\mu=0$ that QL theory does not determine \citep{BeeckWibberenz86}. \citet{Agueda08} use $0.01$ and $0.10$, while \citet{Wijsen19} adopt $0.048$. The mirror interaction fills that gap with an amplitude fixed by the measured compressive power. We define its equivalent constant as $\epsilon_{\rm eq}=\nu(90\arcdeg)/\nu_0$, with $\nu_0$ the gyroresonant rate at $|\mu|\to1$, the measured counterpart of the prescribed constants (Appendix~\ref{app:floor}).

The measured ratio is $0.03$--$0.08$ at 0.5--1~MeV across the two regions, within the range commonly prescribed in transport models, but it varies with energy. It rises as $E^{1.14}$ in the inner precursor and as $E^{0.55}$ in the outer, with the rise concentrated between 1 and 2~MeV, and reaches $0.30\,[0.11,0.84]$ at 5~MeV (Figure~\ref{fig:gap}c). At these energies, the mirror rate grows with particle speed while the receding gyroresonances lower $\nu_0$. Because $\epsilon_{\rm eq}$ is proportional to $\eta_{\rm TTD}$, the factor of ten across the 0.3--3 range propagates directly into $\epsilon_{\rm eq}$, moving the 0.5--1~MeV values from the bottom of the prescribed range at $\eta_{\rm TTD}=0.3$ to its top and beyond at 3. Neither resonance-broadening prescription of \citet{Bieber94}, evaluated on the same measured spectra, reproduces the measured values across 0.5--5~MeV. Dynamical-turbulence damping saturates near 0.1 and falls short above 2~MeV. Random sweeping, the broadening adopted for the gyroresonance itself, stays more than a factor of 50 below throughout, because its wings reach $90\arcdeg$ only from the ion-cyclotron side, where they sample the dissipation-range power at the band top.

\section{The precursor is not steady}
\label{sec:steady}

\begin{figure*}[!t]
  \centering
  \includegraphics[width=0.8\textwidth]{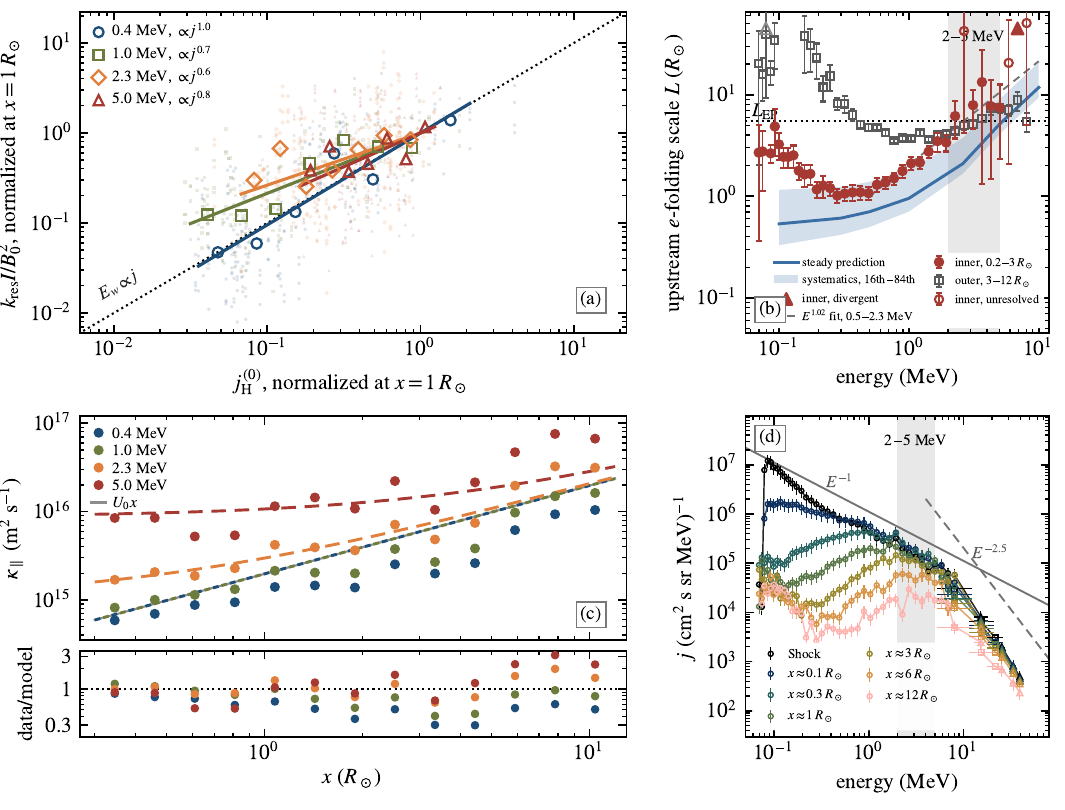}
  \caption{The wave-generation relation, the upstream scales, and the proton spectra of the precursor. \textit{(a)} Resonant slab power $k_{\rm res}I(k_{\rm res})/B_0^2$ against the intensity $j$, both normalized at $x=1\,\Rsun$, for four channels (legend): 15~s values (faint dots), medians in intensity bins (open markers), the fitted $j^{\,g}$ (solid), and the one-to-one relation $g=1$ (dotted). \textit{(b)} $e$-folding scale $L$ in the inner (filled circles) and outer (open squares) windows with $1\sigma$ bars. Open circles mark unresolved inner fits and triangles divergent ones, filled above 2~MeV and open below. The dashed line is the power law of the inner scales, the solid line the scale of the steady profile on the measured $\kpar(E,x)$ at $\eta_{\rm TTD}=1$ with its 16th--84th band, and the dotted line $L_{\rm EP}$. \textit{(c)} Median-binned $\kpar(x)$ at four energies (circles) and the model $\kpar=\kappa_0+U_0x$ fitted on $\kappa_0$ alone (dashed), with $U_0x$ in gray. The strip shows the ratio of the data to each model, dotted at unity. \textit{(d)} Proton spectra at six upstream distances and at the shock from EPI-Lo (circles), LET (squares), and HET (triangles), with the adopted uncertainties as vertical bars and the LET and HET channel widths as horizontal bars. The gray band in (b) and (d) marks the 2--5~MeV transition of Paper~II in the spacecraft frame.}
  \label{fig:confinement}
\end{figure*}

\subsection{The intensity profiles lag the wave growth}

The isotropic proton intensity falls with distance ahead of the shock. We fit its profile against the distance $x$ with an exponential in two windows, $0.2$--$3$ and $3$--$12\,\Rsun$. Convection refreshes the inner window within minutes, so it samples the local wave field, while the outer window retains the accumulated history of the event. In the inner window the measured scales run from $1.1\,\Rsun$ at their minimum, at 0.3--0.5~MeV, to $3.5\,\Rsun$ at 1.7~MeV. The predicted scales are those of the steady profile built on the measured $\kpar(E,x)$, the forward model of Appendix~\ref{app:profiles}. Above 0.5~MeV both the measured and the predicted scales increase with energy (Figure~\ref{fig:confinement}b), but they differ systematically in amplitude. Over 0.4--2.3~MeV the measured scales exceed the predicted ones by a factor of $2.1\pm0.5$ in the channel-by-channel fits and by a factor of $2.0\pm0.3$ in the median-binned profiles, where the spreads are channel-to-channel standard deviations.

The scale ratio, the measured scale over the predicted one, grows from 1.6 at 0.4~MeV to 2.6 at 1.7~MeV, as $E^{0.36\pm0.06}$ on the channel-by-channel fits. Solving the transport equation directly in pitch angle rather than in the diffusion approximation raises the predicted decay rates most in the lowest-energy channels, and with them the scale ratios, to 2.1--2.8. No systematic term reproduces both the magnitude of the ratio and its rise with energy. The calibration, slab-fraction, and phase-speed uncertainties shift the channels nearly together and reach 2.3 in their fully correlated worst case, enough for the overall level but not for its energy dependence. A higher mirror efficiency flattens the rise of the ratio to $E^{0.17}$ at $\eta_{\rm TTD}=3$, but neither it nor a neglected counter-propagating transverse component removes the discrepancy in amplitude. At $\eta_{\rm TTD}=3$ the ratio left to the correlated terms is still 2.7--4.5, above the 2.3 they allow (Appendix~\ref{app:mc}).

The profiles therefore average a coefficient that was larger in the recent past than the instantaneous wave spectra give. The transport equilibrates over $\kpar/u_1^2\simeq4$--$12$~min on the inner medians of these channels, a time that lengthens with energy, so the higher channels average more of the growth and the ratio rises with energy. Because $\kpar$ varies inversely with the resonant wave power, a scale ratio of two requires that power at fixed distance from the shock to have doubled over that time. A single crossing cannot separate this growth in time from the spatial profile, because the observed rise of the resonant wave power contains both.

\subsection{Nonlinear steady balance at one energy at a time}

The resonant wave energy is proportional to the excess of the intensity over its far-upstream value \citep{Bell78,Lee83}, so a self-generated coefficient falls inversely with the intensity it transports, $\kpar=\kappa_0\,j_0/j$ with $\kappa_0$ and $j_0$ their values at the ramp. The diffusion-convection balance then becomes nonlinear in $j$. Where the far-upstream intensity is negligible that balance has a single solution, in which the coefficient grows along the precursor at the closing speed, $\kpar=\kappa_0+U_0x$. Beyond the flattening scale $\kappa_0/U_0$ the intensity falls with power-law index $q=U_0/(d\kpar/dx)=1$ (Appendix~\ref{app:profiles}). We fit neither the growth speed $U_0$ nor the index $q=1$. \citet{Bell78} obtained that $q=1$ profile, while the steady profiles of \citet{Lee83} are exponential in distance, with the power law in energy. Allowing the upstream scattering centers to drift at $v_A$ reduces the predicted growth speed to $(1-v_A/u_1)\,U_0=0.86\,U_0$.

The measured profiles are power laws in distance, $j\propto x^{-q}$ (Figure~\ref{fig:slices}c), as a coefficient that grows linearly with distance produces. Of the two conditions of the steady self-generated solution, the proportionality of the resonant wave power to the local intensity and the growth of $\kpar$ at $U_0$, neither holds across the full energy range. Along the precursor the resonant power $k_{\rm res}I(k_{\rm res})/B_0^2$ at the $|\mu|=1$ resonant wavenumber $k_{\rm res}$ of each channel varies with the isotropic intensity as $j^{\,g}$ (Figure~\ref{fig:confinement}a). The exponent is $g=1.0$ at 0.4~MeV, falling to 0.7 at 1.0~MeV and 0.6 at 2.3~MeV. The proportionality that produces the nonlinear $q=1$ profile therefore holds at the lowest channel but weakens across the confined range. At 5.0~MeV the exponent returns to 0.8, but there it is the ratio of two flattened profiles, since the intensity gradient has flattened to $q=0.55$ and the resonant-power profile flattens with it.

The coefficient does grow with distance, but not at $U_0$. Fixing the growth at $U_0$ and fitting only $\kappa_0$ leaves the binned medians below the model by factors up to 3.4 at 0.4~MeV and 2.5 at 1.0~MeV. At 5.0~MeV the residuals tilt from below unity inside $2\,\Rsun$ to a factor of 3 beyond $5\,\Rsun$ (Figure~\ref{fig:confinement}c). The bin-to-bin intermittency moves the channels together, with a Pearson correlation of 0.8 between the logarithms of the 0.4 and 2.3~MeV estimates within each distance bin, so it cancels in their ratio and does not produce those departures. The fitted $d\kpar/dx$ rises by a factor of $7.8\pm1.3$ from 0.3 to 5~MeV and crosses $U_0$ near 1.3~MeV (Figure~\ref{fig:slices}d). Below that energy $d\kpar/dx<U_0$, so the waves have grown beyond the level the steady balance requires, and above it $d\kpar/dx>U_0$, so the wave growth falls behind. The precursor therefore satisfies the nonlinear steady balance at only one energy at a time.

The measured indices fall with energy, from $q=1.1$--$1.3$ through 1.2~MeV to 0.55 at 5~MeV. Because the steady profile is an exact power law only beyond its flattening scale $\kappa_0/(d\kpar/dx)$, we fit the predicted index to the same steady profile over the same samples (Figure~\ref{fig:slices}c). It runs from 2.3 at 0.4~MeV to 0.9 at 2.3~MeV. At low energy it lies above the measured one by the same factor of two as the ratio of the scales. The two meet at 2.3~MeV, and at 5~MeV the prediction falls below the measurement. On the steady self-generated scaling the number of mean free paths between the shock and any point in the precursor is $x_n/\lpar=v/[3(u_1-v_A)]$, about 2 at 1~MeV. Two mean free paths are too few to establish an asymptotic transport law \citep{Malkov17}, so the index comparison is a consistency check rather than a sharp test.

\section{Confinement fails between 2.3 and 6 MeV}
\label{sec:boundary}

Confinement begins to fail above 2.3~MeV. We measure the $e$-folding scale $L$ of the upstream intensity within $3\,\Rsun$ of the shock, fitted channel by channel (Figure~\ref{fig:confinement}b). Below 2~MeV we resolve the scale to within 21 percent, and it rises with energy to $6.2\pm2.5\,\Rsun$ at 2.3~MeV, reaching the pressure scale $L_{\rm EP}$ that is the depth of the precursor. At 2.7~MeV the intensity does not fall across the window. The fitted decay rate is $0.024\pm0.049$ per $\Rsun$, consistent with zero (the open circle at $42\,\Rsun$), and the same scatter excludes a profile on the $6\,\Rsun$ scale of the channel below at $2.8\sigma$.

Between 3 and 5~MeV the fits return $7$--$13\,\Rsun$ with uncertainties of comparable size, on the $E^{1.0}$ trend of the scales below 2.3~MeV. From 5.9~MeV upward we resolve no scale, because a scale above $25\,\Rsun$ changes the intensity by about a tenth across the $2.8\,\Rsun$ window, less than the scatter of the samples. The release is therefore a transition, with its onset at 2.3~MeV, where the upstream scale reaches the depth of the precursor, and its end near 6~MeV, where the gradient vanishes.

The energy at which the gradient first vanishes shifts between 2.3 and 5.9~MeV with the fit window (Appendix~\ref{app:profiles}), and we measure the transition from the particle profiles alone. Paper~II locates the same transition without any upstream gradient, from the reversal of the downstream anisotropy at $2$--$5$~MeV in the spacecraft frame and above 3~MeV in the shock frame. The measured onset lies inside that band. The released particles still cross the precursor, and Paper~II identifies them as the weakly scattered population in a wave field whose amplitude the lower-energy protons set. The outer window measures their gradient and above 2.7~MeV returns a scale of $5$--$9\,\Rsun$, straddling $L_{\rm EP}$ (Figure~\ref{fig:confinement}b), so even after the inner gradient vanishes their intensity still falls across the whole precursor on its pressure scale. Over the transition the resonant wave field stops confining particles within the precursor, and the transition lies below the maximum energy of the accelerated spectrum.

Below the $2$--$5$~MeV band the $E^{-1}$ spectrum of the shock-accelerated population survives only at the shock itself (Paper~II; \citealp{Jebaraj24}). Across the band every upstream spectrum steepens toward the escaping slope, reaching $E^{-2.5}$ over 5--45~MeV inside $3\,\Rsun$ and flattening to $E^{-2.2}$ and $E^{-2.1}$ at the two outermost distances (Figure~\ref{fig:confinement}d). The shock spectrum keeps its $E^{-1}$ form through the band although the precursor does not confine these particles. The amplitude measured at the ramp implies a coefficient there near the Bohm limit, well below the precursor value, so particles at the onset still return to the front and the shock spectrum steepens only at higher energy. Above the band the particles observed upstream do not return to the front to gain energy \citep{PtuskinZirakashvili05,Caprioli10,Malkov13}, so their spectrum freezes at escape.

Three limits apply to the energy the shock confines, and the two that follow from $\kpar$ alone, the age of the shock and the size of the accelerator, both lie above the onset. Reaching energy $E$ takes $t_{\rm acc}=\int(20\kpar/u_1^2)\,d\ln p$, with $p$ the proton momentum (\citealp{Drury83}; \citealp{MalkovDrury01}), for a compression ratio of 4 and equal coefficients across the front, integrated upward from 0.05~MeV. On the $\kpar(E)$ of the last half solar radius the integral is 69 minutes at 2.6~MeV. With the coefficient continued above 3~MeV along the measured $E^{0.86}$ rise of the mean free path, $\kpar\propto E^{1.36}$, it meets the 4~h age of the shock at $\simeq$8.9~MeV.

The size limit requires the diffusion length to fit inside the accelerator, and the near-ramp $\kpar/U_0$ reaches the depth of the precursor $L_{\rm EP}$ at $\simeq$9.5~MeV, so the onset lies a factor of four below both limits. The depth of the precursor is not an independent size, however. On the coefficient implied by the profiles themselves, $U_0L$ with $L$ the fitted inner scale, the diffusion length is $L$ itself, so it reaches $L_{\rm EP}$ at the onset. An independent size is the focusing length of the diverging magnetic field, $26\,\Rsun$ for the Parker field at 0.24~au \citep{Vainio14}. The diffusion length in the frame of the scattering centers, $\kpar/(u_1-v_A)$, reaches that scale only at $\simeq$24~MeV, ten times the onset.

Both limits inherit the factor of two by which the measured profile scales exceed those predicted from the wave-derived coefficient. Doubling the coefficient to $2.0\,\kpar$ lowers the age limit to 4.8~MeV and the limit set by the depth of the precursor to 5.7~MeV, a factor of two above the onset and inside the transition. The independent focusing limit falls to 14~MeV. The remaining terms move that margin downward to 1.3 or upward by up to a factor of three. Converting distances with $U_0$ instead of $u_1$ raises the energy at which the gradient first vanishes from 2.7 to 3.7~MeV and leaves the corrected limits, evaluated with $u_1$, a factor of 1.3 to 1.5 above it (Appendix~\ref{app:profiles}). 

We do not measure the downstream coefficient. The compressed downstream turbulence implies a downstream value $\kappa_2$ below the upstream $\kpar$ (Paper~II), which shortens $t_{\rm acc}$ toward the limiting upstream contribution $4\kpar/u_1^2$ and raises the energy the age allows by up to a factor of four. Evaluating the upstream residence on the growing $\kpar(x)$ rather than its near-ramp value lowers that energy by 1--4 percent.

The third limit, the growth time of the waves, does not follow from $\kpar$. Here $\kpar$ is a property of the waves the particles have already made, and making them takes time. Paper~II infers that time from the evolution of the folding energy, the energy at which each upstream spectrum turns over into its power law, and finds it reaching $\tau_{\rm c}$ at 2.4~MeV, at the onset of the transition. The folding energy at a distance $x$ is also the energy whose upstream scale equals $x$. The 2.4~MeV of Paper~II and the onset are therefore the same feature of the intensity, one found along energy and the other along distance. The scale ratio of Section~\ref{sec:steady} places the coefficient the particles experienced above the one derived from the instantaneous wave spectra at every channel through the onset. The confining waves were therefore still growing when the profiles formed.

\section{Discussion and conclusions}
\label{sec:discussion}

\begin{figure}[!t]
  \centering
  \includegraphics[width=\columnwidth]{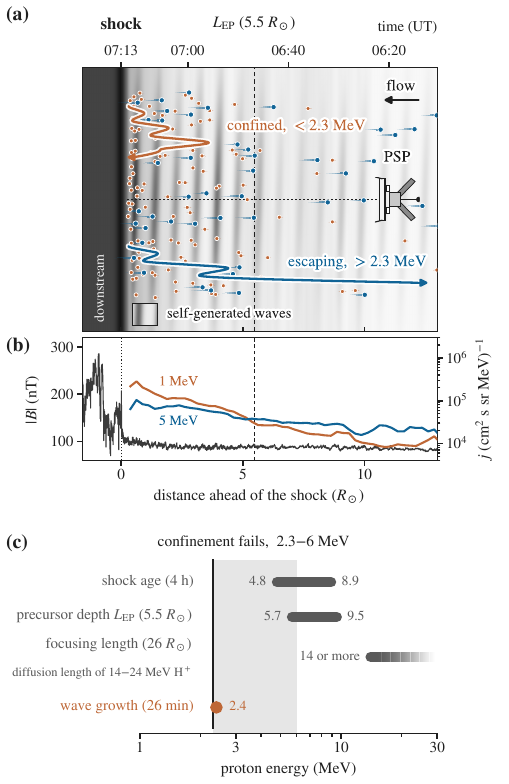}
  \caption{The precursor and the limits on the energy it confines. \textit{(a)} Schematic of the precursor on the distance axis of (b), with a confined proton below 2.3~MeV and an escaping proton above it. \textit{(b)} Field strength and proton intensities. \textit{(c)} The confinement boundary against the three limits, bars spanning the doubled and the measured coefficient.}
  \label{fig:summary}
\end{figure}

We have measured the diffusion coefficient of a shock from its own waves, in energy and in distance, and tested the steady self-generated solution against the particles it transports. The mean free path grows with distance as $x^{0.79}$--$x^{0.90}$, close to the linear rise of the steady solution, and matches its amplitude at 1~MeV. Near that energy, the mirror force of the compressive fluctuations produces a minimum in $\lpar$ that the magnetostatic scaling misses. The growth speed of the coefficient, however, equals the steady-state value $U_0$ only near 1.3~MeV. The intensity profiles are a factor of two shallower than the measured coefficient predicts at every channel through the onset of the transition, because the waves are still growing. Confinement fails between 2.3 and 6~MeV, where the inner scale of the intensity reaches the depth of the precursor and then the gradient vanishes (Figure~\ref{fig:summary}).

The waves reach the steady level at one energy at a time. Below the crossing they lie above the level the wave-generation relation requires, because the intensity there sits at the streaming-limited plateau \citep{ReamesNg10} that Paper~II identifies for this event. At the crossing they meet that level, above it they fall short, and beyond the onset they stop confining the particles, while at every energy through the onset they are still growing. Simulations that retain the pitch angle in the resonance condition show the same continued growth, because protons of higher energy keep driving waves that also scatter those of lower energy \citep{Afanasiev15,Afanasiev23}. Each energy therefore reaches the DSA steady state at its own time. The confinement boundary is the energy at which the growth time of the waves equals the residence time $\tau_{\rm c}$.

Neither the age of the shock nor its size sets the energy it confines. Both limits follow from $\kpar$ and rise with the speed of the shock, and they lie a factor of two to four above the confinement boundary (Figure~\ref{fig:summary}c). The third limit, the growth of the waves, instead depends on the particles that generate them. The waves at each energy grow only while streaming protons drive them, and a plasma element spends only the residence time $L_{\rm EP}/u_1$ in the precursor for the integrated streaming to reach the growth threshold \citep{Vainio03}. A faster shock raises the age and the size limits but shortens the residence time, so its particles must grow the waves faster to confine what those limits allow. At this shock they have not, and Paper~II infers the growth reaching the residence time at the onset.

Above the confinement boundary the $E^{-1}$ spectrum of the shock survives only at the ramp, where the Bohm-limited layer still returns the particles at the onset, and every upstream spectrum has steepened to the released $E^{-2.5}$. The released particles still fall off across the whole precursor on its pressure scale (Figure~\ref{fig:summary}b). The break of the upstream spectrum is the boundary itself, the folding energy that Paper~II tracks. The shock spectrum continues past the boundary to a rollover above $\simeq$10--20~MeV (Paper~II), and the event to proton rigidities above 1~GV ($\simeq$430~MeV) at Earth \citep{Dresing25}.

A traveling shock evolves throughout its life, so the confinement boundary does not stay at one energy. As the waves grow, the boundary moves to higher energy and the break of the escaping spectrum moves with it. The particles that leave are those just above the boundary, so an observer connected to the shock would see the higher energies arrive later. Recent events show that order at their onset, the inverse velocity arrival \citep{Chen25,Li25,Kouloumvakos26b,Xu26}, which the studies attribute to acceleration still in progress at an initially weak shock or to evolving connectivity. The growth of the confining waves provides a physical mechanism for the former: the maximum energy released from the shock rises as sufficient resonant wave power is built at progressively higher rigidity. Simultaneous particle and wave measurements can test that mechanism directly.

A supernova remnant has a boundary of the same kind, set by the time the escaping particles need to grow the field, and it determines which energy escapes at each age \citep{Bell13}. At any given time, the escaping particles are those at the instantaneous maximum energy, so the released spectrum follows the time history of that energy rather than the spectrum at the shock \citep{PtuskinZirakashvili05,Caprioli10}. With self-generated magnetic-field amplification, that energy peaks near the onset of the Sedov stage and falls thereafter \citep{Caprioli10,Bell13}.

PSP crossed this shock four hours after its eruption, while the waves were still growing and the confinement boundary lay well below the energies permitted by the shock's age and size. A single crossing shows the boundary at one moment and cannot tell how fast it moves to higher energy. The speed of the boundary decides the highest energy the shock confines in its lifetime. As the shock expands and its field and scattering environment change, the boundary need not rise monotonically, and the limits set by age and size move as well. Those two limits are the standard estimates of the highest energy a shock can reach, from coronal shocks to supernova remnants. Does any traveling shock ever confine the energies that its age and its size allow, or does the growth of its own waves set the limit at every age?

\section*{acknowledgements}
I.C.J. and E.H. were supported by the Research Council of Finland (X-Scale, grant No.~371569).
I.C.J. was also supported by FWO senior grant No.~1295826N.
I.C.J. acknowledges support from ISSI's ``Visiting Scientist Program''.
M.M. was supported by NASA under the ATP awards 80NSSC24K0774 (UCSD) and 80NSSC26K0593 (Eureka Sci.) and the Fermi award 80NSSC25K7346. 
N.W. acknowledges funding from projects G024226N (Research Foundation Flanders) and 3E241013 (KU Leuven starting grant).  
The work of O.V.A. was supported by NASA grants 80NSSC22K0521, 80NSSC21K1690, 80NSSC23K1278, and NSF GEM grant 2555054.
A.K. acknowledges financial support from the NASA contract NNN06AA01C (Parker Solar Probe EPI-Lo) and from NASA’s LWS grant 80NSSC25K0130.
The work of F.F. was partially supported by NASA grant 80NSSC21K0119 (Living With a Star Program).
A.A. and R.V. acknowledge support by the Research Council of Finland (ENERGISE, grant No. 364853; SpaceResilience, grant
No. 374097).

\facilities{PSP (FIELDS, SWEAP, IS$\odot$IS)}

\software{\texttt{numpy} \citep{Harris20}, \texttt{scipy} \citep{Virtanen20},
\texttt{matplotlib} \citep{Hunter07}, \texttt{cdflib}, \texttt{PyWavelets}}

\appendix

\section{From the wave field to the transport coefficients}
\label{app:method}

\subsection{Calibrated wave spectra}

We resample the MAG field \citep{Bale16}, 293~Hz from 04:00~UT to the ramp, to 12.5~Hz and decompose it in a local field-aligned frame built from a 60~s background field, with $\hat{z}\parallel\bm{B}_0$, $\hat{x}$ in the $\bm{B}_0$--flow plane, and $\hat{y}$ completing the triad. We high-pass all three components with the same 60~s window, so the parallel channel excludes the slow rise of $|B|$ toward the ramp. We rescale the complex Morlet wavelet spectra of the three components, on 72 logarithmic frequencies between 0.0062 and 6~Hz, per frequency so that their time-averaged power matches the Welch spectral density of the same signal. We evaluate both on the record ending 375~s before the ramp.

We divide out two deterministic responses. The first is the high-pass response $|1-\mathcal{D}(f)|^2$, with $\mathcal{D}$ the Dirichlet kernel of the 60~s window. The second is the removal of slow rotations by the rotating basis, measured per frequency as the ratio of fixed-basis to rotating-basis Welch spectra and reaching a factor of 2.4 near 8~mHz. We mask the six band-bottom frequencies, up to 10~mHz, where the combined removal leaves less than 15 percent of the power. We average the powers within each 15~s bin with the cone of influence masked, which gives $P_{xx}$, $P_{yy}$, and $P_{zz}$ and the transverse power $P_\perp=P_{xx}+P_{yy}$. The Welch spectrum of the full $3$--$20\,\Rsun$ record, formed with a slab fraction of 0.90 (Figure~\ref{fig:gap}a), lies a factor of 2 to 4 above the 15~s medians at their low-wavenumber end, more than the intermittency accounts for. That end is the $|\mu|=1$ resonance of $\simeq$50~MeV protons, twenty times the energies at the onset of the transition.

\subsection{Slab fraction and wavenumber conversion}

The slab fraction of the transverse power follows from the reduced-spectrum ratio $\rho=P_{yy}/P_{xx}$ of the two-component model \citep{Bieber96}, inverted after aggregation as $f_{\rm slab}=2(s_K-\rho)/[2(s_K-\rho)+(1+s_K)(\rho-1)]$ with $s_K=5/3$ the index of the two-dimensional component, the slab component giving $\rho=1$ for any spectrum. The inversion requires $\rho\in[1,s_K]$, so we clip the ratios below 1, at 38 percent of the frequencies and corresponding to slab power polarized in the $\bm{B}_0$--flow plane, to 1, and interpolate over the 8 percent above $s_K$. The median fraction over the 66 analyzed frequencies is 0.83, and over the 36 inside the domain of the inversion it is 0.60. It is a share of the reduced power, which at the near-parallel field-to-flow angles of this interval weights slab above two-dimensional power, so it exceeds the fluctuation-energy fractions of PSP two-component decompositions \citep{LZhao22,Bandyopadhyay21}. The slab waves are not axisymmetric about $\bm{B}_0$, so the separation is approximate and the fraction enters through the systematic range 0.6--1.0 of Appendix~\ref{app:mc}, which spans the in-domain and the clipped values. The wave-vector statistics of Paper~I's transverse families, whose occurrence peaks below $15\arcdeg$ with medians of $16\arcdeg$--$23\arcdeg$ and a tail to large angles, set the lower edge of that range.

The spacecraft-frame frequency converts to parallel wavenumber through $k=2\pi f/V_\varphi$, with $V_\varphi=V_{\rm adv}+v_A$ and $V_{\rm adv}$ the advection speed projected on the field, equal to the normal flow $V_n$ to one percent at this obliquity. The one-sided spectrum is $I(k)=f_{\rm slab}(P_{xx}+P_{yy})\,V_\varphi/2\pi$ (Figure~\ref{fig:gap}a). The conversion is Alfv\'enic at every $k$, and it matches the branch phase speeds to better than 8 percent at the gyroresonant wavenumbers, $kd_i\lesssim0.3$. We quote the band-top values of Section~\ref{sec:coeff} on this grid, and a branch-consistent assignment would move the right-handed branch top from $kd_i\simeq1.9$ to $\simeq$1.4 and $\omega_R/\Omega$ from 4.5 to 2.6, with the tangency inside the band on either assignment. The driven band of each region begins at the first wavenumber at which the local spectral slope flattens to within $\pm0.5$ and ends at the break of a two-segment power-law fit. We fit the falls quoted in Section~\ref{sec:coeff} from the break to $kd_i\simeq0.3$, and fits from twice the break to $kd_i=1$ steepen to $-2.3$ in the outer region and $-2.7$ in the inner.

We retain bins only while the field-to-flow angle stays within $30\arcdeg$, so that the reduced spectrum samples quasi-parallel wavenumbers, which excludes 05:55--06:12~UT (Figure~\ref{fig:shock}a). The spacecraft-potential electron density, calibrated against quasi-thermal noise, gives the densities and Alfv\'en speed quoted in Section~\ref{sec:coeff}, and the proton temperature from the Solar Probe Analyzer for Ions \citep[SPAN-Ai;][]{Kasper16,Livi22} gives the sound speed that enters the fast speed $c_f$.

\subsection{The scattering coefficients}

Because the branches propagate in a single, anti-sunward direction along the field, the gyroresonant part of the pitch-angle diffusion coefficient is \citep{NgReames95,Vainio00}

\begin{align}
D_{\mu\mu}={}&
\frac{\pi}{2}\,\frac{\Omega^2}{B_0^2}\,(1-\mu^2)\notag\\
&\times\sum_{n=\pm1}\int I_n(k)\,
R_\sigma\!\left[\omega_n(k)-kv\mu-n\Omega\right]dk,
\label{eq:dmumu}
\end{align}

with $\mu$ the pitch-angle cosine in the solar-wind frame, measured along the wave-propagation direction, $v$ the proton speed in the same frame, $\Omega$ its relativistic gyrofrequency, and $I_{\pm1}=I(k)(1\mp\sigma_m)/2$ the branch weights set by the measured reduced helicity, positive for the right-handed branch about $\bm{B}_0$. The frequency $\omega_n(k)$ is the cold-plasma parallel dispersion of the resonant branch, written $\omega_R$ for $n=-1$ and $\omega_L$ for $n=+1$ and evaluated from the background field and the Alfv\'en speed, and $R_\sigma$ is the random-sweeping resonance function, a unit-area Gaussian of frequency width $\sigma_\omega=\alpha v_Ak$. This width follows from the random-sweeping correlation of \citet{Bieber94} written as $\exp(-\alpha^2v_A^2k^2t^2/2)$, whereas their $\exp(-\alpha^2v_A^2k^2t^2)$ gives $\sqrt{2}\,\alpha v_Ak$. We set the free parameter $\alpha$ to the measured slab amplitude $\delta B_\perp/B_0$ of each interval, integrated over its full measured band.

A kinetic Maxwellian dispersion at the measured proton beta $\beta_p=0.20$ leaves the whistler branch undamped and within one percent of the cold root through the measured band, with the tangency shifted by 0.3 percent. It damps the ion-cyclotron branch at $\gamma/\omega=0.11$ at $kd_i=0.8$, and above $kd_i\simeq1.3$ the damping exceeds half the real frequency. The Maxwellian omits the energetic protons that drive these waves, so that damping is an upper limit. The ion-cyclotron edge therefore lies between $\mu=-0.002$ on the cold branch and $\mu=-0.014$ with the branch cut at $\gamma/\omega=0.1$ on the cold real frequency, and the gap is 2.6--3.0 times the non-dispersive width across that range. The non-dispersive comparison evaluates the same spectra with $\omega=kv_A$ and the matching resonance Jacobian. The three dispersion treatments of Section~\ref{sec:coeff} are these evaluations on the cold branches, with the damping cut, and non-dispersive.

The mirror interaction with the compressive family adds to Equation~(\ref{eq:dmumu}) the transit-time damping term

\begin{equation}
    D_{\rm TTD}=\eta_{\rm TTD}\,\frac{\pi}{4}\,\frac{v}{B_0^2}\,\left(1-\mu^2\right)^2\,G_w\left(\mu-\mu_c\right)\int\! k_\parallel\,S_\parallel(k_\parallel)\left[\frac{2J_1(\xi)}{\xi}\right]^2 dk_\parallel,
\label{eq:dttd}
\end{equation}

with $\mu_c=c_f(\theta_{kB})/(v\cos\theta_{kB})$ the pitch-angle cosine of the resonant speed of Section~\ref{sec:coeff}, $G_w$ a unit-area Gaussian of standard deviation $w$, $S_\parallel(k_\parallel)$ the measured spectrum of the compressive family, and $J_1$ the first-order Bessel function of $\xi=k_\perp v_\perp/\Omega$. The factor $[2J_1(\xi)/\xi]^2$ is the finite-gyroradius response of the mirror force \citep{Schekochihin2009,YanLazarian04,Xu18}, and it tends to unity as $\xi\to0$, where the integral reduces to $\langle k_\parallel\rangle\,\delta B_\parallel^2$. On this field the factor averaged over the compressive band, with the transverse spectral shape as weight, is 0.96 at 0.4~MeV and 0.77 at 2.6~MeV, and 0.97 and 0.81 with the measured compressive shape, within 2 percent in $\lpar$. 

On unperturbed orbits $G_w$ is a delta function, which would leave $D_{\mu\mu}=0$ on the rest of the gap and $\lpar$ infinite. Summed over wave normal angles, linear transit-time damping still requires $v\mu\geq v_A$ \citep{Schlickeiser98} and never reaches $\mu=0$. The mirror force spreads the parallel speed of a resonant proton by $\Delta v_\parallel=v_\perp(\delta B_\parallel^2/B_0^2)^{1/4}$, which broadens the resonance at every wavenumber into $G_w$ with $w=\Delta v_\parallel/v=(1-\mu_c^2)^{1/2}(\delta B_\parallel^2/B_0^2)^{1/4}$ and the same area \citep{Volk75,Xu18}. The resonance lies inside the gap, at $\mu_c=0.17$, 0.099, and 0.054 for 0.1, 0.3, and 1~MeV protons in the inner precursor. The median of $w$ rises from 0.11 in the outer precursor to 0.15 in the inner, against a gap 0.066 wide at 1~MeV, so in the median interval $D_{\rm TTD}$ varies by less than 12 percent over the gap at that energy. At 0.1~MeV it falls at the ion-cyclotron edge to a quarter of its peak in the outer precursor and to half in the inner.

The integral of \citet{HasselmannWibberenz70} then gives the mean free path,
\begin{equation}
\lpar=\frac{3v}{8}\int_{-1}^{1}\frac{(1-\mu^2)^2}{D_{\mu\mu}}\,d\mu.
\label{eq:lambda}
\end{equation}

\subsection{Branch weights and the numerical evaluation}

\begin{figure}[!t]
  \centering
  \includegraphics[width=0.5\columnwidth]{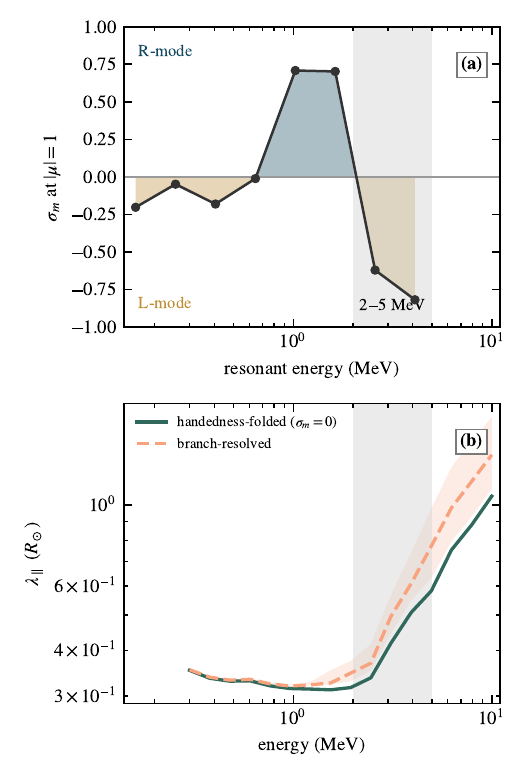}
  \caption{Handedness of the wave field against energy, and the near-shock mean free path with and without it. \textit{(a)} Reduced helicity $\sigma_m$ within $1\,\Rsun$ of the ramp against the energy of the proton whose $|\mu|=1$ gyroresonance falls at each wavenumber bin, shaded blue for $\sigma_m>0$ (right-handed branch) and tan for $\sigma_m<0$ (left-handed branch). \textit{(b)} $\lpar(E)$ within $1\,\Rsun$ of the ramp, with equal branch weights (solid) and branch-resolved (dashed), the band spanning $\sigma_m$ tables of 11, 23, and 47 bins. The gray band spans the 2--5~MeV transition of Paper~II in the spacecraft frame.}
  \label{fig:branch}
\end{figure}

The branch weights come from Paper~I's per-cell signed ellipticity in its $\bm{B}_0$-referenced convention, rebinned as the reduced helicity $\sigma_m(k)$ on 23 wavenumber bins and applied at all distances (Figure~\ref{fig:branch}a). Within $1\,\Rsun$ of the ramp $\sigma_m(k)$ reaches $|\sigma_m|\simeq0.6$--$0.8$ at the resonant wavenumbers of 1--4~MeV protons and reverses sign between them, so $D_{\mu\mu}$ is asymmetric in $\mu$ (Figure~\ref{fig:gap}b). An evaluation with equal branch weights ($\sigma_m=0$) moves $\lpar$ by under 5 percent below 1.5~MeV, by 10--18 percent at the 2--2.7~MeV channels, and by up to 33 percent at 5--10~MeV. The resonances of the 2--2.7~MeV channels fall on the sign reversal, where the spread across tables of 11--47 wavenumber bins reaches 4--33 percent. The equal-weight evaluation shifts the acceleration-time and diffusion-length limits by about 13 percent (Figure~\ref{fig:branch}b).

The mirror interaction takes $\delta B_\parallel^2$ as a third of $P_{zz}$ integrated above 10~mHz, the fast-magnetosonic share of the parallel power, and $\langle k_\parallel\rangle$ as the power-weighted mean parallel wavenumber of that family over $9\times10^{-5}$--$3\times10^{-4}$~km$^{-1}$. The spread of the family's wave normal angles over $50\arcdeg$--$76\arcdeg$ moves $\mu_c$ between 0.04 and 0.11 at 1~MeV, within the broadening width. The in-phase amplitude of Paper~I, whose band begins at 3~mHz, is $0.017\,B_0$ in the outer precursor and $0.025\,B_0$ in the inner, and the mirror rate on it is higher by at most a third, because the rate weights the power by wavenumber. The share scales the wavenumber integral of Equation~(\ref{eq:dttd}) linearly and the broadening width as its fourth root. Any share from a ninth of $P_{zz}$ to all of it keeps $\lpar$ at 0.3--2.6~MeV and $\epsilon_{\rm eq}$ at 0.5--5~MeV inside the efficiency range of Appendix~\ref{app:mc}. At 0.05~MeV beyond $0.5\,\Rsun$ a ninth lengthens $\lpar$ by factors of 3.5 to over 4, against 2.6--2.9 at $\eta_{\rm TTD}=0.3$.

We evaluate Equation~(\ref{eq:lambda}) on 401 pitch-angle nodes with no spectral extrapolation beyond the measured band. At the measured $\alpha=0.07$--$0.26$ the resonance integral reduces to the resonance-point form $I_n(k_{\rm res})/|d\omega_n/dk-v\mu|$ weighted by the Gaussian mass inside the band, and agrees with a direct numerical integral on $4\times10^5$ wavenumber nodes to 4 percent above 1~MeV and 20 percent at 0.05~MeV. We locate each root at the sign change of $\omega_n(k)-kv\mu-n\Omega$ on a 300-node logarithmic wavenumber grid over the band, interpolate it with $I_n$ and $d\omega_n/dk$ in $\ln k$, and take its Gaussian mass as the error-function difference between the band edges. Inside the dispersive gap the condition has no root and the resonance-point form gives zero. The Gaussian tail of the direct integral enters the gap only from the ion-cyclotron side and stays below 2 percent of the mirror term at $90\arcdeg$ at 1~MeV and 9 percent at 0.1~MeV in the inner precursor.

The omitted wave-frame factor $(1-\mu\omega/kv)^2$ deviates from unity by under 5 percent at 1~MeV and 20 percent at 0.05~MeV, below unity on the right-handed branch and above it on the left-handed one. Its net effect on $\lpar$ is therefore smaller than those values and cancels where the branches balance. The corresponding factor of the mirror term, $(1-\mu\mu_c)^2$, equals unity at $\mu=0$ and leaves $\epsilon_{\rm eq}$ unchanged, and it raises $\lpar$ by 0.1--0.2 percent at 1~MeV and 2--5 percent at 0.05~MeV.

We reject estimates without an in-band gyroresonance in both hemispheres at $|\mu|=0.95$, whose resonances cover less than half the pitch-angle integral, or that return $\lpar>50\,\Rsun$. The fixed-distance curves of Figure~\ref{fig:slices}a are medians over the 15~s bins within a factor of 1.35 of each distance, widened to 1.6 where fewer than two bins fall inside and truncated at $13\,\Rsun$. The $\lpar(x)$ statistics are medians in 41 logarithmic distance bins over $0.08$--$13\,\Rsun$, the joint slope is an unweighted fit of the logarithmic medians over $0.3$--$13\,\Rsun$, and the amplitude at $x=1\,\Rsun$ and 1~MeV has an 8 percent standard error.

\section{Particle profiles and the forward model}
\label{app:profiles}

\subsection{Intensity, spectra, and profile fits}

The monopole $j^{(0)}_{\rm H}$ is the EPI-Lo proton intensity of the ChanP data product averaged over eight uniform $\mu$ bins of its 80 apertures and then across those bins. That average cancels the first- and second-order anisotropies, including the field-aligned beam and the Compton-Getting dipole, to corrections of order $(V_n/v)^2$. We keep zero-count look directions, since masking them biases the medians of each channel upward by factors of 2 to 40 over the fit window.

The spectra of Figure~\ref{fig:confinement}d combine EPI-Lo with the Low Energy Telescope (LET) and High Energy Telescope (HET) of IS$\odot$IS \citep{McComas16,Hill17}. We sample each upstream spectrum at the time when the shock stood at the stated distance, with 1-minute LET and HET integrations after 07:00~UT, taken at the nearest column within $0.3\,\Rsun$ of that distance. Before 07:00~UT the integrations are 15 minutes long, each spanning $3.1\,\Rsun$ of shock travel. The shock spectrum is the average from the ramp to 3.5 minutes behind it for EPI-Lo, and of the 07:13 and 07:14~UT minutes for LET and HET. We adopt a 30 percent intensity uncertainty for EPI-Lo and HET.

We fit the profiles as $\ln j$ against $x$ by least squares in the inner ($0.2$--$3\,\Rsun$) and outer ($3$--$12\,\Rsun$) windows for the 38 EPI-Lo channels with valid data. The inner window holds 13 samples and the outer 44 at the 60~s cadence, excluding any zero-intensity samples. We call a fit unresolved, and the gradient vanished, when its scale exceeds $25\,\Rsun$, when its uncertainty exceeds 1.5 times the scale, or when the slope is non-negative and the fit diverges. We do not draw outer fits with scales above $45\,\Rsun$ or errors above 1.5 times the scale in Figure~\ref{fig:confinement}b. The inner window leaves five channels unresolved, at 0.079, 2.658, 5.865, 6.867 and 8.063~MeV. The 0.079~MeV case is in the suprathermal background rather than at the confinement boundary, and we draw it with an open marker.

The point estimates are $L=6.15\,\Rsun$ at 2.3~MeV, a slope of $-0.163$ per $\Rsun$, and $42.4\pm88.0\,\Rsun$ at 2.7~MeV, a slope of $-0.024\pm0.049$ on samples with an rms scatter of 0.14 in $\ln j$. A nonparametric bootstrap with a fixed seed, $2\times10^4$ draws resampling the 13 profile samples of the two channels independently, returns a longer scale at 2.7~MeV than at 2.3~MeV in 91 percent of draws. A divergent fit, 25 percent of the 2.7~MeV draws and 4 percent of the 2.3~MeV draws, counts as longer than any finite scale, and over the draws in which both fits converge the fraction is 93 percent. The window variants of Section~\ref{sec:boundary} contract the inner window by half a solar radius at one edge or extend its outer edge to $3.5$--$4.0\,\Rsun$. The contraction keeps the energy at which the gradient first vanishes between 2.3 and 3.7~MeV, and the extension lengthens the fitted scale before it vanishes and moves that energy to 5.9~MeV.

The power-law index is a least-squares fit of $\ln j$ against $\ln x$ on identical samples, 57 per channel over $0.2$--$12\,\Rsun$, applied to the data and, at the same sample positions, to the forward-modeled profile. Of the two estimates of the scale ratio in Section~\ref{sec:steady}, the first compares the channel-by-channel fits with the forward model at the same energies. The second fits the same exponential to median-binned profiles on a logarithmic grid in $x$ of 5 to 12 bins, for which the ratio runs from 1.9 to 2.1.

Below 0.2~MeV the measured profile flattens, and below $\simeq$0.13~MeV the fitted scale rises to $2$--$5\,\Rsun$, a factor of 5 to 9 above the prediction, where the transport is not diffusive (Section~\ref{sec:coeff}) and the steady balance does not apply. These channels contain a suprathermal population related to reflected ions \citep{Jebaraj24}, and background-subtracted refits move the fitted scale by less than 25 percent.

\subsection{Validity of the steady balance}

The front approaches at $U_0$ while the distance conversion uses $u_1$, so the two differ by $U_0/u_1=1.18$. This factor cancels in every fitted exponent and in $t_{\rm acc}$, and enters the diffusion-length limit through $\kappa_{\rm eff}=U_0L_{\rm EP}$ and the residence time through $\tau_{\rm c}=L_{\rm EP}/u_1$. The prediction of Figure~\ref{fig:confinement}b is the steady diffusion-convection profile $j_{\rm mod}(x)\propto\exp[-\int_0^{x}U_0\,dx'/\kpar(x')]$ \citep{Drury83}, built for each energy on the measured $\kpar(x)$ and fitted over the same inner window in the same way as the data, on the model's 15~s grid. The band is a 400-member Monte Carlo, with the calibration drawn log-normally at a $1\sigma$ factor of 1.4, the phase-speed conversion normally at 8 percent, the slab fraction uniformly over 0.6--1.0, and $\eta_{\rm TTD}$ log-uniformly over 0.3--3. The calibration and phase-speed terms enter as exact scale factors, and we propagate the slab fraction and the mirror efficiency, which do not, through full forward models at three values each, interpolated in the logarithm. The prediction rises monotonically with energy and therefore does not reproduce the turnover of the measured scales below 0.5~MeV.

The steady balance is the small-anisotropy limit of the pitch-angle Fokker-Planck equation, and the streaming anisotropy $3u_1/v$ reaches 0.5 at 1~MeV and unity at $\simeq$0.3~MeV (Section~\ref{sec:coeff}). We solve the exact steady eigenvalue problem $\partial_\mu(D_{\mu\mu}\partial_\mu h)=-\Lambda(v\mu-u_1)h$ for the coefficient of Equations~(\ref{eq:dmumu}) and (\ref{eq:dttd}), with $h(\mu)$ the pitch-angle eigenfunction and $\Lambda$ the spatial decay rate in the shock frame, in a Legendre basis converged at 600 modes. The exact solution raises the predicted decay rate by 25 percent at 0.4~MeV and 8--9 percent at 1--2.3~MeV, which raises the scale ratio of Section~\ref{sec:steady}. Above 4~MeV it lowers the rate by 2--3 percent.

The diffusion approximation also requires the intensity gradient scale to exceed $\lpar$. At 1~MeV $\lpar$ is $0.12$--$0.56$ of the fitted inner scale across $0.3$--$3\,\Rsun$ and $0.95$--$2.1$ of the fitted outer scale beyond $6\,\Rsun$, so the approximation holds through the inner window and fails in the outermost bins of the $\lpar(x)$ fit of Section~\ref{sec:coeff}. The balance applies in the convectively refreshed inner window, so we show the outer window without a model curve. A time-dependent solution with the measured, static $\kpar(x)$ reproduces the steady inner-window scales at the 4~h age of the shock. The focusing length of Section~\ref{sec:boundary} leaves the fitted inner gradient flatter by under a tenth below 2~MeV, and the adiabatic momentum loss over one residence time is of order one percent.

We neglect transport across the field, which the wandering of the field lines controls \citep{JokipiiParker69}. The integral of the measured transverse correlation to its first zero, in the mean-field frame and summed over the two transverse components, gives a field-line diffusion coefficient $D_{\rm FL}=0.03$--$0.07\,\Rsun$ on the pre-event wind and $0.006$--$0.02\,\Rsun$ on the $3$--$20\,\Rsun$ record. The range spans the detrending and the wave-frame conversion. On the larger value the parallel walk $(2\kpar\tau_{\rm c})^{1/2}$ carries a particle $[2D_{\rm FL}(2\kpar\tau_{\rm c})^{1/2}]^{1/2}=0.4$--$0.9\,\Rsun$ across the field in one $\tau_{\rm c}$ over the 0.4--2.6~MeV channels. That displacement is an effective ratio of cross-field to parallel diffusion, $\kappa_\perp/\kpar$, of $0.01$--$0.05$, which brackets the $0.02$--$0.04$ of test-particle simulations in three-dimensional turbulence at interplanetary amplitudes \citep{GiacaloneJokipii99}. The displacement is a fraction of the inner window, so the fitted gradient is the gradient along the sampled field line to that accuracy.

\subsection{The wave-generation relation}

With the resonant wave energy proportional to the excess intensity, written in the negligible-background limit as $\kpar=\kappa_0 j_0/j$, the steady balance $\kpar\,\partial j/\partial x+U_0 j=U_0 j_\infty$ becomes nonlinear and integrates to
\begin{equation}
j(x)=j_\infty\left[1-\left(1-\frac{j_\infty}{j_0}\right)e^{-j_\infty U_0 x/j_0\kappa_0}\right]^{-1},
\label{eq:nlprofile}
\end{equation}
with $j_\infty$ the intensity far upstream. Along this solution $d\kpar/dx=U_0(1-j_\infty/j)$, so for $j\gg j_\infty$ the coefficient is $\kpar=\kappa_0+U_0x$ and the profile is $j=j_0/(1+U_0x/\kappa_0)$, the $q=1$ power law of Section~\ref{sec:steady}. The channel-by-channel proportionality is the local, steep-spectrum form of the growth-rate integral of \citet{Bell78} and \citet{Lee83}.

The growth speed $d\kpar/dx$ of Figure~\ref{fig:slices}d is a linear fit to the median-binned $\kpar(x)$ over $0.3$--$12\,\Rsun$, with 16th--84th intervals from a bootstrap over the distance bins. The same fit restricted to the inner window puts the crossing at 2.0~MeV, and the $v_A$ drift of Section~\ref{sec:steady} lowers it from 1.3 to 1.1~MeV. We fit the $\kappa_0$ of Figure~\ref{fig:confinement}c on the same binned medians by least squares in $\ln\kpar$ with $\kappa_0$ constrained positive, which returns $1.0\times10^{15}$~m$^2$\,s$^{-1}$ at 2.3~MeV and $8.7\times10^{15}$~m$^2$\,s$^{-1}$ at 5.0~MeV. At 0.4 and 1.0~MeV the fit runs to $\kappa_0=0$, so no member of the family lies low enough to reach the data. The resonant-power exponent $g$ is the ratio of the fitted spatial log-gradients of $k_{\rm res}I(k_{\rm res})/B_0^2$, at the magnetostatic $|\mu|=1$ resonance of each channel, and of the isotropic intensity over the same $0.2$--$12\,\Rsun$. The residual 15~s fluctuations about the two profiles have Pearson correlations of $-0.2$ to $+0.3$ across the channels, so the fitted exponent is a property of the profiles rather than of the bin-to-bin variations.

\section{Systematics}
\label{app:mc}

Four systematic factors add to the 16th--84th intermittency spreads defined in Section~\ref{sec:coeff}. The absolute spectral calibration contributes a factor of 1.4 and the phase-speed conversion 8 percent. The slab fraction spans $0.6$--$1.0$ against the measured 0.83, with $\lpar$ scaling inversely where gyroresonance dominates the integral. The mirror efficiency $\eta_{\rm TTD}\in[0.3,3]$ rescales $\lpar$ over the last half solar radius by factors of $0.49$--$2.6$ at 0.3~MeV, narrowing to $0.84$--$1.5$ at 2.6~MeV. An efficiency at either end of the range breaks the amplitude match of Section~\ref{sec:coeff}.

The mirroring-broadened width of \citet{Volk75} and \citet{Xu18}, applied to the same spectra in place of the random-sweeping width of Equation~(\ref{eq:dmumu}), shortens $\lpar$ by factors of $4.8$, $3.1$, and $1.8$ at $0.4$, $1.0$, and $2.6$~MeV near the ramp. It flattens the energy dependence of the scale ratio and raises it to 4--6. That width is the $\mu\to0$ limit of the parallel-velocity excursion produced by the compressive fluctuations, which falls as $v_\perp^2\delta B_\parallel/(2v_\parallel B_0)$ away from $90\arcdeg$, so the factors above are upper limits on the change.

Two further terms are one-sided. A counter-propagating transverse fraction of 5--10 percent, below what the dispersion analysis resolves, scatters the hemisphere otherwise left to the mirror and lowers $\lpar$ at 0.1~MeV by a factor of 1.4--1.7. It leaves the boundary energies, the energies of the transition of Section~\ref{sec:boundary}, unchanged. Equation~(\ref{eq:dttd}) assigns the compressive power to a single propagation sense. A field with both senses moves half the power to a $-\mu_c$ lobe outside the gap and raises $\lpar$ by a third at 0.1~MeV and 3 percent at 1~MeV. With the whole power sunward, $\lpar$ inside $3\,\Rsun$ rises by a factor of 3 to 5 at 0.1~MeV, 1.3 to 1.4 at 0.3~MeV, and 6 percent at 1~MeV. Beyond $3\,\Rsun$ it rises by a factor of 1.7 at 0.3~MeV and 9 percent at 1~MeV, and exceeds $50\,\Rsun$ at 0.1~MeV in most intervals. At the boundary energies it rises by 1--3 percent. Near the ramp its rise by an order of magnitude at 0.05~MeV lengthens $t_{\rm acc}$ at 2.6~MeV from 69 to 115 minutes, and energies below 0.1~MeV add three quarters of the increase. On the coefficient scaled to match the profiles in this case, $1.9\,\kpar$, the age limit falls to $\simeq$3~MeV.

In quadrature, the calibration, slab-fraction, phase-speed, and mirror-efficiency terms span a factor of 1.5 downward and 2.0 upward at the boundary energies, and even the fully correlated worst case reaches 3.5, or 2.3 without the mirror term. The minimum of $\lpar(E)$ survives the mirror efficiency and the counter-propagating fraction, though its depth does not, and the two move its location over 0.5--1.6~MeV at fixed distance. \citet{Jebaraj24} measure the shock speed as $u_1=2400\pm300$~km\,s$^{-1}$, and that uncertainty moves the distance coordinate and the steady prediction together. Over its range the fitted amplitude at $x=1\,\Rsun$ and 1~MeV runs from 1.24 times the predicted one at 2100~km\,s$^{-1}$ to 0.79 times it at 2700~km\,s$^{-1}$, so the amplitude comparison of Section~\ref{sec:coeff} holds to 20--25 percent.

\section{The equivalent 90-degree constant}
\label{app:floor}

The comparison of the mirror interaction of Equation~(\ref{eq:dttd}) with the prescribed constants of Section~\ref{sec:coeff} uses the scattering rate $\nu(\mu)=2D_{\mu\mu}(\mu)/(1-\mu^2)$. The prescribed forms are $\nu=\nu_0\,(|\mu|^{s-1}+H)$ \citep{BeeckWibberenz86} and $\nu=\nu_0\,(|\mu|/(1+|\mu|)+\epsilon)$ \citep{Agueda08,AguedaVainio13}, in which $s$ is the spectral index of the resonant slab spectrum and $H$ or $\epsilon$ the rate at $90\arcdeg$ in units of $\nu_0$. In the measured ratio the numerator is the mirror rate of Equation~(\ref{eq:dttd}) at $\mu\to0$, $\nu(90\arcdeg)=2D_{\rm TTD}(\mu\to0)$, whose amplitude and width are both measured up to the efficiency $\eta_{\rm TTD}$. The denominator is the hemisphere-symmetrized $|\mu|\to1$ limit of the gyroresonant rate with the mirror term off, $\nu_0=[D_{\mu\mu}(0.99)+D_{\mu\mu}(-0.99)]/(1-0.99^2)$, which removes the sensitivity to the sign convention of $\sigma_m$. This $\nu_0$ is the normalization of \citet{BeeckWibberenz86}, and in the \citet{Agueda08} convention the equivalent constant is $\epsilon_{\rm eq}/2$. We compute the ratio $\epsilon_{\rm eq}(E)=\nu(90\arcdeg)/\nu_0$ per 15~s interval wherever the measured band resolves both rates, and the slopes of Figure~\ref{fig:gap}c are power-law fits to the region medians over 0.5--5~MeV.

We evaluate the two resonance-broadening models of \citet{Bieber94} on the measured field (Figure~\ref{fig:gap}c). In the dynamical-turbulence damping model the fluctuations decorrelate exponentially at the rate $\alpha v_Ak$, on the same measured $\alpha$, so the resonance acquires Lorentzian wings that reach $90\arcdeg$. The random-sweeping function of Equation~(\ref{eq:dmumu}) decorrelates as a Gaussian in time, and its exponentially suppressed wings lie eighteen broadening widths from $90\arcdeg$ on the whistler side of the gap in the inner precursor. On the ion-cyclotron side the gap edge lies within one width, but the wings there sample only the dissipation-range power at the band top, so the gyroresonant rate at $90\arcdeg$ stays more than an order of magnitude below the mirror rate.

\bibliography{transport}
\bibliographystyle{aasjournal}

\end{document}